\documentclass[aps,pre,twocolumn,amsmath,amsfonts,amssymb,amsthm,superscriptaddress,showpacs]{revtex4-1}

\usepackage{epsfig,color}

\newcommand{\eq}[1]{Eq.~(\ref{eq:#1})}

\newcommand{\Eq}[1]{Equation~(\ref{eq:#1})}
\newcommand{\fig}[1]{Fig.~\ref{fig:#1}}

\newcommand{\app}[1]{appendix~\ref{app:#1}}

\newtheorem{theorem}{Theorem}

\begin{document} 

\title{Effects of sampling interaction partners and competitors in evolutionary games}

\author{Christoph Hauert}
\affiliation{Department of Mathematics, The University of British Columbia, 1984 Mathematics Road, Vancouver, B.C., Canada, V6T 1Z2}

\author{Jacek Mi{\c{e}}kisz}
\affiliation{Institute of Applied Mathematics and Mechanics, University of Warsaw, Banacha 2, 02-097 Warsaw, Poland}

\date{ \today}



\begin{abstract}
The sampling of interaction partners depends on often implicit modelling assumptions, yet has marked effects on the dynamics in evolutionary games. One particularly important aspect is whether or not competitors also interact. Population structures naturally affect sampling such that in a microscopic interpretation of the replicator dynamics in well-mixed populations competing individuals do not interact but do interact in structured populations. In social dilemmas interactions with competitors invariably inhibit cooperation while limited local interactions in structured populations support cooperation by reducing exploitation through cluster formation. These antagonistic effects of population structures on cooperation affect interpretations and the conclusions depend on the details of the comparison. For example, in the  Snowdrift game, spatial structure may inhibit cooperation when compared to the replicator dynamics. However, modifying the replicator dynamics to include interactions between competitors lowers the equilibrium frequency of cooperators, which changes the conclusions and space is invariably beneficial, just as in the Prisoner's Dilemma. These conclusions are confirmed by comparisons with random matching models, which mimic population structures but randomly reshuffle individuals to inhibit spatial correlations. Finally, the differences in the dynamics with and without interactions among competing individuals underlies the differences between death-birth and birth-death updating in the spatial Moran process: death-birth supports cooperation because competitors tend not to interact whereas they tend to do for birth-death updating and hence cooperators provide direct support to competitors to their own detriment.
\end{abstract}

\maketitle 

\section{Introduction}
Cooperation between unrelated individuals in animal and human societies is an intriguing issue in biology and social sciences \citep{hamilton:AmNat:1963,axelrod:book:1984,hammerstein:book:2003,sigmund:book:2010,nowak:book:2006}. In the Darwinian world of \emph{survival of the fittest}, altruistic behaviour that is costly to the individual and benefits others should be selected against, yet cooperation is ubiquitous in nature. Cooperation represents a social dilemma because everyone is better off cooperating but each individual is tempted to free-ride on the benefits created by others, which creates a conflict of interest between the group and the individual \citep{dawes:ARP:1980}. This fundamental challenge in behavioural sciences can be addressed with evolutionary game theory where the Prisoner's Dilemma, the Snowdrift game or the Stag-hunt game represent different instances and variants of the underlying social dilemma \citep{hauert:JTB:2006a}.

In the Prisoner's Dilemma cooperators pay a cost, $c$, to provide a benefit, $b$, to their interaction partner ($b>c$), while defectors neither incur costs nor provide benefits. If two cooperators meet they mutually benefit from the interaction but both participants face the temptation to defect and avoid the costs of cooperation. However, if both participants choose to defect neither one gains anything from this interaction. The Prisoner's Dilemma represents to most stringent form of a social dilemma. In order to maintain cooperative behaviour, positive assortment among cooperators is required \citep{fletcher:PRSB:2009}, i.e. cooperators must  more likely interact with other cooperators than in random encounters. The necessary assortment of cooperators, or, more precisely, of acts of cooperation can be achieved through various mechanisms, including conditional response in repeated interactions through direct or indirect reciprocity \citep{trivers:QRB:1971,nowak:Nature:1993,alexander:book:1987,nowak:Nature:1998} or through limited local interactions in spatially structured populations \citep{ohtsuki:Nature:2006,szabo:PR:2007}. In particular, population structures impose constraints on the sampling of interaction partners. These constraints not only affect an individual's fitness by restricting interactions to a subset of the entire population but also limit its exposure to individuals with alternative strategies that could be imitated or whose offspring might displace the individual. More specifically, in the Prisoner's Dilemma, population structures facilitate assortment by enabling cooperators to form clusters and thereby reducing exploitation from defection.

The Snowdrift game relaxes the social dilemma to the extent that cooperators and defectors can co-exist in the absence of assortment \citep{sugden:book:1986}.
The name of the game refers to the situation where two drivers on their way home are trapped on either side of a snowdrift. If both cooperate and start shovelling they both get the benefit of getting home while splitting the costs for clearing the way, $b-c/2$. However, if only one shovels both still get the benefit but the cooperator bears the full costs. Finally if no one shovels no one gets anywhere. In contrast to the Prisoner's Dilemma, the best strategy now depends on the opponent: if the other cooperates it is better to defect, as before, but if the other defects it is better to cooperate and get $b-c$ instead of nothing for mutual defection. Note that the Snowdrift game turns into a Prisoner's Dilemma for large costs, $c>b$, and for even larger costs, $c>2b$, cooperation is no longer a viable strategy, the dilemma disappears and defection becomes the mutually preferred strategy.

Formally, the Snowdrift game is equivalent to the chicken or Hawk-Dove game 
\citep{maynard-smith:Nature:1973} but provides an interpretation in terms of cooperation rather than conflict and competition. Interestingly, the effects of assortment and spatial structure, in particular, on cooperation in the Snowdrift game \citep{hauert:Nature:2004} are not as clear-cut as in the Prisoner's Dilemma. In particular, the conclusion whether spatial structure promotes or inhibits cooperation depends on the details of the reference setup and ultimately on the sampling of interaction partners and competitors. 

Finally, the Stag-hunt game is an instance of a coordination game where the best strategy also depends on the opponent but this time it is best to do the same as the opponent. The Stag-hunt game is inspired by Rousseau's social contract \citep{rousseau:book:1755} noting that hunting a stag requires the concerted cooperative efforts of a hunting party but each hunter faces the temptation defect and catch a hare instead. Although the hare feeds the hunter, doing so spoils the group's efforts to bag the preferred stag. In this case, the conflict of interest reduces to a coordination game because even though both players prefer mutual cooperation, they may get trapped in states of mutual defection because neither party has an incentive to (unilaterally) switch strategy.

Here we consider the effects of sampling individuals for interaction and competition on the evolutionary dynamics and, in particular, on cooperation in social dilemmas. More specifically, we consider sampling in well-mixed, i.e. unstructured populations, for the classical, deterministic replicator dynamics \citep{hofbauer:CUP:1998,taylor:MB:1978} as well as stochastic finite population models \citep{woelfing:JTB:2009,traulsen:PRL:2005} and compare the dynamics to random-matching models \citep{robson:JET:1996b,miekisz:JTB:2005,molzon:DGA:2012} as well as structured population models \citep{nowak:Nature:1992b,szabo:PR:2007}.

\section{Sampling in infinite populations}
The replicator dynamics \citep{hofbauer:CUP:1998,taylor:MB:1978,sandholm:book:2010} describes evolutionary changes in infinite populations that consist of $d$ strategic types with frequencies $x_i$ for $i=1,\ldots,d$:
\begin{align}
\label{eq:rep}
\dot{x_i} =\ & x_i(f_i-\bar f),
\end{align}
where $f_i$ denotes the average fitness of type $i$ derived from random interactions with other members of the unstructured (well-mixed) population and $\bar f=\sum_{i=1}^d x_i f_i$ indicates the average fitness of the population. Naturally $\sum_{i=1}^d x_i=1$ must hold and hence the dynamics unfolds on the simplex $S_d$. \Eq{rep} states that any strategic type, which performs better than the population on average, increases in abundance. More specifically, the fitness $f_i$ reflects the payoffs achieved in interactions with other members of the population. If interactions occur randomly and among pairs of individuals (as opposed to larger groups), the payoffs are given by a matrix ${\bf A} = [a_{ij}]$, where the element $a_{ij}$ indicates the payoff of an individual of type $i$ interacting with an individual of type $j$. In that case the replicator equation can be written in matrix form
\begin{align}
\label{eq:repmatrix}
\dot{x_i} =\ & x_i(({\bf A}{\bf x})_i-{\bf x}{\bf A}{\bf x}),
\end{align}
with $f_i=({\bf A}{\bf x})_i$, $\bar f={\bf x}{\bf A}{\bf x}$ and ${\bf x}=(x_1,\ldots,x_d)$ denoting the current state of the population.

From a microscopic perspective, the evolutionary dynamics unfolds in two stages: First, all individuals interact with other members of the population by playing games and receiving payoffs according to the payoff matrix $\bf A$, which then affects and determines their fitness. Second, individuals compete for reproduction and succeed at rates (or probabilities) proportional to their fitness, which changes the composition of the population, $\bf x$. Together the two stages result in a Darwinian process based on variation (differences in traits) and selection (differences in fitness). In a cultural context, the population composition changes if individuals preferentially imitate or learn strategies of more successful types, i.e. those that have a higher fitness $f_i$. Similarly, in a biological (genetic) context, the population composition changes through differential reproduction rates, again represented by the fitness $f_i$, where offspring replace other members of the population. The replicator equation, \eq{rep}, represents the infinite population limit of either process \citep{traulsen:PRL:2005,traulsen:PRE:2006a,traulsen:bookchapter:2009}.

\begin{figure*}[tb]
\centerline{\includegraphics[width=\linewidth]{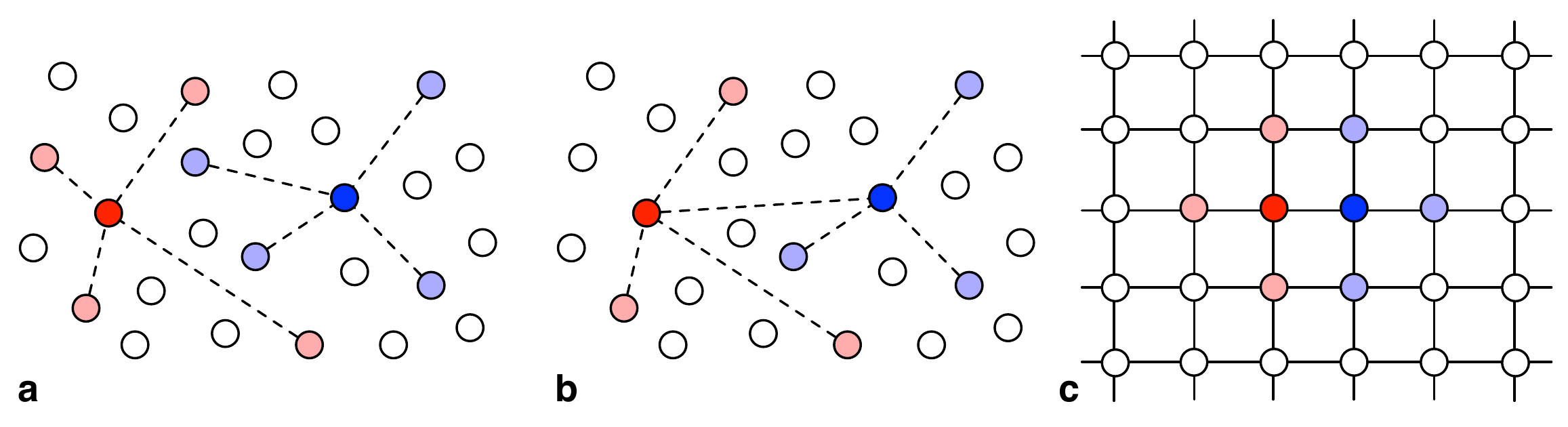}}
\caption{\label{fig:sampling}Sampling of interaction partners in well-mixed (\textbf{a} and \textbf{b}) and spatially structured populations (\textbf{c}). \textbf{a} The two competitors (red and blue) each interact with a random sample of $k=4$ other members of the population (light red and light blue, respectively). \textbf{b} The two competitors interact with each other plus $k-1=3$ random members of the population. \textbf{c} In spatially structured populations competitors are always neighbours and interact with each other as well as their other neighbours, some of which may be shared depending on the underlying geometry.
}
\end{figure*}

The fitness $f_i$ in \eq{rep} is determined by the payoffs of an individual of type $i$ when interacting with other members of the population in state $\bf x$. If each individual interacts with $k$ randomly chosen partners, then there are, on average, $k x_i$ interaction partners of type $i$, see \fig{sampling}a. For $f_i$ based on average payoffs, $f_i=({\bf A}(k{\bf x}))_i/k=({\bf A}{\bf x})_i$, the matrix form \eq{repmatrix} is recovered and the average fitness remains unaffected by the number of interactions $k$. Only the variance of the fitness between individuals of the same type increases for small $k$ and vanishes for $k\to\infty$ but in infinite populations this has no effect on the evolutionary dynamics.

Implicit in the conclusion that the dynamic is independent of $k$ is the assumption that the focal and model individual do not interact and hence assumes that no correlation exists between interactions and competition. However, if the two competitors, the focal and model individual, \emph{do} interact and only the remaining $(k-1)$ interaction partners are randomly sampled, then a focal individual of type $i$ and a model of type $j$ have, on average, fitness
\begin{subequations}
\label{eq:fifj}
\begin{align}
f_i&=\frac{(k-1)({\bf A}({\bf x}))_i+a_{ij}}{k}\\
f_j&=\frac{(k-1)({\bf A}({\bf x}))_j+a_{ji}}{k},
\end{align}
\end{subequations}
respectively, see \fig{sampling}b. Clearly, the effects of sampling on fitness vanish in the limit $k\rightarrow\infty$ and are strongest for small $k$, such that for $k=1$ individuals exclusively interact with their current competitor. 

The replicator dynamics \eq{rep} is recovered for pairwise comparison processes where a focal individual $u$ compares its payoff to a model individual $v$ \cite{traulsen:PRL:2005}. In a cultural context, the focal individual adopts the model's strategy at a rate proportional to their payoff difference $f_v-f_u$, whereas, in a genetic context, the focal individual gets displaced by (clonal) offspring of the model. Suppose $p_{i\to j}$ is the rate at which an individual of type $i$ adopts the strategy of (or gets displaced by) an individual of type $j$. Then the rate of change of the frequency of type $i$ is given by
\begin{align}
\label{eq:pij}
\dot x_i &= \sum_{j=1}^d x_i x_j p_{j\to i}-\sum_{j=1}^d x_i x_j p_{i\to j}\nonumber\\
	&= x_i \sum_{j=1}^d x_j\left(p_{j\to i}-p_{i\to j}\right).
\end{align}
If the rates $p_{j\to i}$ are linear functions of the payoff differences, $f_i-f_j$, then \eq{pij} recovers the replicator dynamics, \eq{rep}. For example, this applies for the pairwise comparison process given by 
\begin{align}
\label{eq:pji}
p_{j\to i}=\frac12+\frac{\omega}2(f_i-f_j),
\end{align}
where $\omega\geq0$ indicates the selection strength. For $\omega=0$ selection is absent and the process is neutral, whereas for large $\omega$ even small fitness differences translate to large selective advantages. $\omega\ll1$ refers to the important limit of weak selection. This yields $\dot x_i=\omega x_i(f_i-\bar f)$, which is the same as \eq{rep} except for a constant rescaling of time by $\omega$. Hence the net effect of changing the selection strength is a change in the time scale of the replicator dynamics. Often it is useful to interpret $p_{j\to i}$ as transition probabilities in which case an upper bound, $\omega_{\text{max}}$, exists to ensure that $p_{j\to i}$ remains confined to $[0,1]$.

\subsection{$2\times2$ games}
In order to highlight the differences in dynamics due to the sampling process, consider pairwise interactions, two strategies, $A$ and $B$, and the generic, symmetric payoff matrix
\begin{align}
\label{eq:2x2}
\bordermatrix{
  & \mathrm{A} & \mathrm{B} \cr
\mathrm{A} & \alpha & \beta \cr
\mathrm{B} & \gamma & \delta \cr}.
\end{align}
The replicator equation, \eq{rep}, then reduces to
\begin{align}
\label{eq:repab}
\dot x &= x\left(f_A-\bar f\right) = x(1-x)\left(f_A-f_B\right)
\end{align} 
where $x$ denotes the frequency of type $A$ individuals. The dynamics admits two trivial equilibria at $x=0$ and $x=1$ plus possibly a third, interior equilibrium, $x^\ast$, for which $f_A=f_B$ holds. The existence and location of $x^\ast$ depends on the sampling process to select interaction partners for type $A$ and $B$ individuals, which then determines $f_A$ and $f_B$, respectively. If all $k$ interaction partners are randomly sampled the interior equilibrium is given by
\begin{equation}
\label{eq:eqint}
x^\ast=\frac{\delta-\beta}{\alpha-\beta-\gamma+\delta},
\end{equation}
provided that $0<x^\ast<1$ and is independent of $k$. Moreover, the interior equilibrium is invariant to adding constants to columns in the payoff matrix, which is a consequence of the corresponding invariance of the replicator dynamics, \eq{rep}. Consequently the generic dynamics can be effectively reduced to two instead of four parameters (payoffs) \citep{hauert:IJBC:2002} but this generally does not extend to other dynamics or population structures. Naturally, \eq{eqint} also represents the mixed Nash equilibrium of the game, \eq{2x2}. However, if competing individuals always interact (see \eq{fifj}) then, for $k\geq2$, the interior equilibrium is shifted to
\begin{equation}
\label{eq:eqintk}
x^\ast_k=x^\ast-\frac1{k-1}\frac{\beta-\gamma}{\alpha-\beta-\gamma+\delta}.
\end{equation}
The shift solely originates in the choice of sampling scheme to determine the fitness of $A$ and $B$ type individuals. Moreover, the interior equilibrium, $x^\ast_k$, no longer preserves the invariance of $x^\ast$ and hence the generic dynamics now relies on three payoffs (one can still be absorbed in a rescaling of time). In the limit $k\to\infty$ the differences vanish and $x^\ast_k\to x^\ast$. For $k=1$ the fitness of an individual is exclusively determined by the interaction with its competitor and the replicator equation (\ref{eq:repab}) reduces to
\begin{align}
\dot x &= x(1-x)\left(\beta-\gamma\right)
\end{align} 
and hence no interior equilibrium exists regardless of the game. In fact, any game is reduced to a purely competitive interaction -- if $A$ outperforms $B$ ($\beta>\gamma$) then $A$-types keep increasing and $x=1$ is the only stable state. The converse holds if $B$ outperforms $A$ ($\beta<\gamma$) and $A$-types dwindle and disappear.

In the following we focus on canonical forms of social dilemmas involving only one or two parameters (payoffs) while maintaining the characteristic and representative features of the interaction.

\subsubsection{Prisoner's Dilemma}
The Prisoner's Dilemma is characterized by the payoff ranking $\gamma>\alpha>\delta>\beta$, which ensures that $B$ (defection) dominates $A$ (cooperation) and the third equilibrium $x^\ast$ does not exist. The donation game \citep{sigmund:book:2010} outlined in the introduction is a popular instance of the Prisoner's Dilemma
\begin{align}
\label{eq:pd}
\bordermatrix{
  & \mathrm{C} & \mathrm{D} \cr
\mathrm{C} & b-c & -c \cr
\mathrm{D} & b & 0 \cr},
\end{align}
which can be re-scaled and reduced to a single parameter interaction based on the cost-to-benefit ration $r=c/b$. Moreover, the donation game satisfies equal-gains-from-switching \citep{nowak:AAM:1990}, $\gamma-\alpha=\beta-\delta$, and hence renders the replicator dynamics frequency independent. The modified sampling scheme yields
\begin{align}
\dot x &= x(1-x)\left(-c-\frac1k b\right).
\end{align} 
Changing the sampling scheme only rescales time but does not affect the outcome -- cooperators remain doomed and actually disappear even faster for decreasing $k$.

\subsubsection{Snowdrift game}
\begin{figure}[tp]
\centerline{\includegraphics[width=0.8\linewidth]{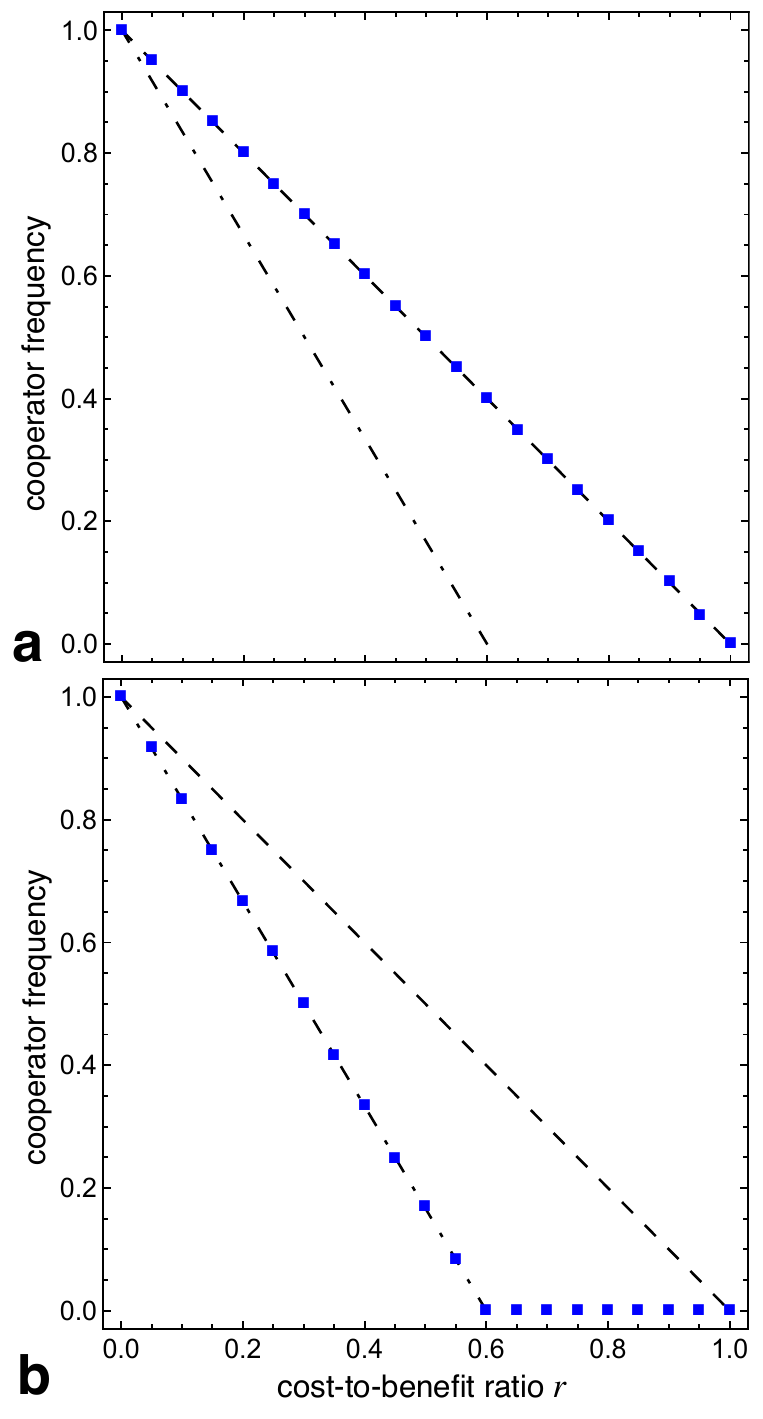}}
\caption{\label{fig:sdmix}Equilibrium fraction of cooperators in well-mixed populations interacting in the Snowdrift game, $x^\ast$ (dashed line) and $x^\ast_k$ (dash-dotted line) for $k=4$, as a function of the cost-to-benefit ratio $r$ compared to individual based simulations (\textcolor{blue}{\scriptsize$\blacksquare$}) for $N=10^4$ and selection strength $\omega=0.01$.
\textbf{a} random sampling of interaction partners matches $x^\ast$ even close to $0$ or $1$ indicating that for $N=10^4$ stochastic fluctuations, which may result in the extinction or fixation of cooperators, are negligible. 
\textbf{b} modified sampling scheme to include the competitor among the interaction partners matches $x^\ast_k$ and is consistently lower than $x^\ast$. Above $r_c=(k-1)/(k+1)=3/5$ cooperators can no longer persist.
}
\end{figure}

The characteristic payoff ranking of the Snowdrift game is very similar with $\gamma>\alpha>\beta>\delta$, i.e. only the last inequality is reversed, which renders $A$ (cooperation) attractive when facing $B$ (defection) and consequentially admits a globally stable interior fixed point $x^\ast$. Traditionally the Snowdrift game is parametrized as
\begin{align}
\label{eq:sd}
\bordermatrix{
  & \mathrm{C} & \mathrm{D} \cr
\mathrm{C} & b-\frac c2 & b-c \cr
\mathrm{D} & b & 0 \cr},
\end{align}
with $b>c>0$ such that $x^\ast=1-r$ with the cost-to-benefit ratio of mutual cooperation $r=c/(2b-c)$. Thus, if cooperation is cheap, $c\ll b$, most individuals cooperate at equilibrium, whereas if it is expensive, $c\lesssim b$, few cooperators persist. Regardless of costs and benefits, the equilibrium state is always a stable mixture of cooperators and defectors. This is no longer the case when changing the sampling scheme to always include the competing individual. For $k\geq2$ this shifts the equilibrium to 
\begin{align}
\label{eq:xksd}
x^\ast_k = 1-\frac{k+1}{k-1}r = x^\ast-\frac2{k-1}r
\end{align}
and hence invariably lowers the fraction of cooperators at equilibrium. In particular, this change introduces a threshold $r_c=(k-1)/(k+1)$ above which cooperation is no longer sustainable, see \fig{sdmix}a~\&~b.

\subsubsection{\label{sect:sh}Stag-hunt game}
The Stag-hunt game is an instance of a coordination game characterized by the payoff ranking $\alpha>\gamma \geq \delta>\beta$ such that it is always best to choose the same strategy as the opponent. The traditional parametrization of the Stag-hunt game is given by the payoff matrix:
\begin{align}
\label{eq:sh}
\bordermatrix{
  & \mathrm{C} & \mathrm{D} \cr
\mathrm{C} & 1 & 0 \cr
\mathrm{D} & a & a \cr},
\end{align}
where the value of a stag is conveniently normalized to $1$ and a hare is worth $1>a>0$. Similar to the Snowdrift game, the Stag-hunt game also admits an interior fixed point $x^\ast=a$ but now $x^\ast$ is unstable and the two trivial equilibria $x=0$ and $x=1$ are both stable. Consequently, $x^\ast$ also marks the basin of attraction for each of the two stable, homogeneous equilibria. In populations with $x>x^\ast$ the frequency of cooperators keeps increasing while decreasing if $x<x^\ast$. If $x^\ast>1/2$ then defection is risk dominant (has the larger basin of attraction) and if $x^\ast<1/2$ cooperation is risk dominant but in either case mutual cooperation is the preferred, or efficient, outcome \citep{harsanyi:book:1988,kandori:ECO:1993}. Hence cooperation is risk dominant as long as hares are lean and worth less than half a stag, $a<1/2$.

Changing the sampling scheme to always include the competing individual shifts the separating equilibrium to 
\begin{align}
\label{eq:shxk}
x^\ast_k = \frac{k}{k-1}a = x^\ast+\frac1{k-1}a
\end{align}
for $k\geq2$ and hence consistently increases the basin of attraction of defection. For $k(1-a)\leq1$ (including $k=1$) the basin of attraction of efficient cooperation disappears and defection remains the sole globally stable equilibrium.

\subsection{Structured populations}
In structured populations individuals do not randomly interact with any other member of the population but rather interactions are confined to an individual's neighbourhood as defined by a graph structure where each vertex represents an individual and edges reflect interactions. Consequently, the fitness of individuals is stochastic and based on the particular realization of the strategies in their neighbourhood rather than expected payoffs as in \eq{rep} or \eq{pij}.

\begin{figure}[tp]
\centerline{\includegraphics[width=0.8\linewidth]{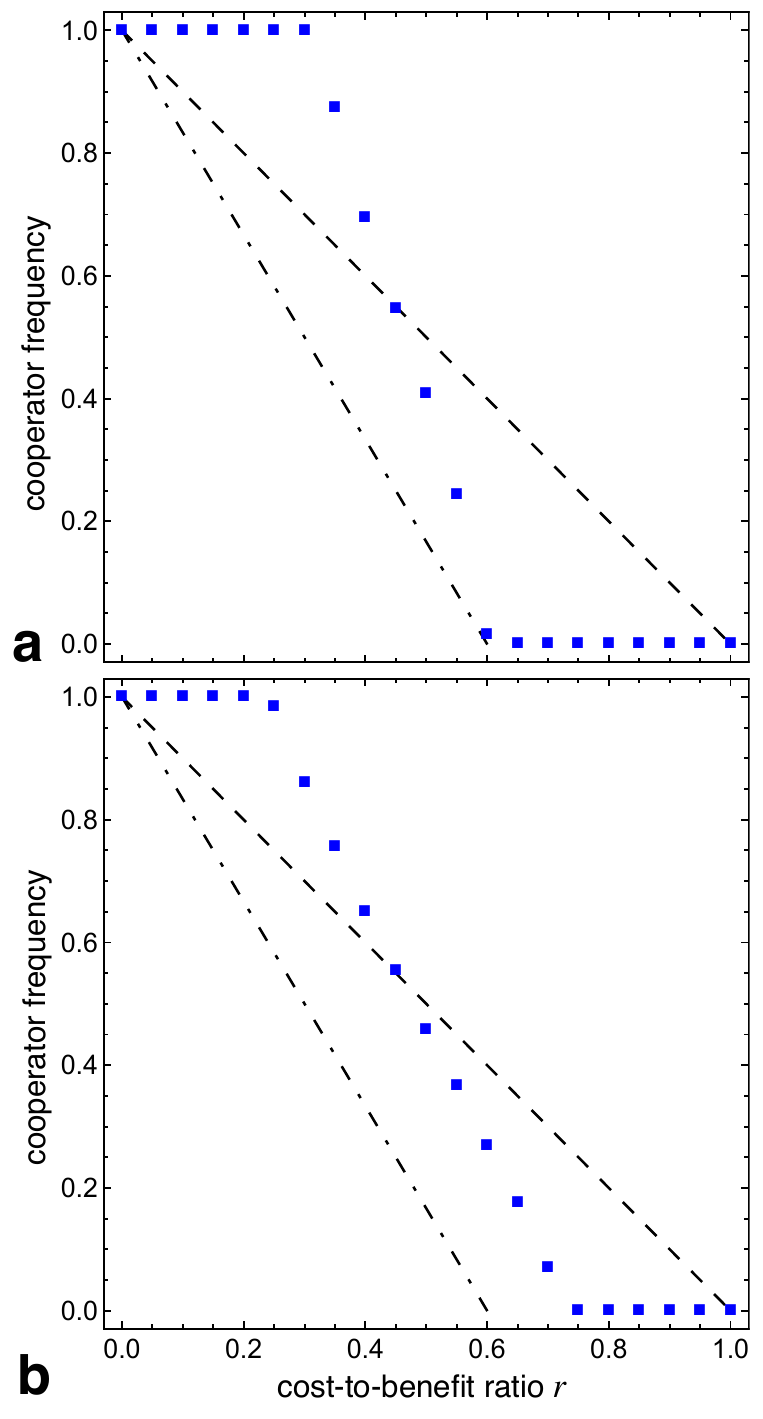}}
\caption{\label{fig:sdstruct}Equilibrium fraction of cooperators in structured populations interacting in the Snowdrift game as a function of the cost-to-benefit ratio $r$ from individual based simulations (\textcolor{blue}{\scriptsize$\blacksquare$}) with selection strength $\omega=0.01$. As a reference the equilibrium of the replicator dynamics is shown for the standard sampling scheme, $x^\ast$ (dashed line), as well as for the modified sampling scheme $x^\ast_k$ (dash-dotted line) with $k=4$.
\textbf{a} lattice population ($N=100\times100$) interacting with the four nearest neighbours ($k=4$). Compared to $x^\ast$, spatial structure turns out to be detrimental to cooperation for larger $r$, however, compared to $x^\ast_k$ spatial structure remains beneficial. 
\textbf{b} random regular graph with four neighbours ($N=10^4, k=4$) results in equilibrium frequencies of cooperators between the lattice and $x^\ast$ due to the decrease in ordered structure and hence local clustering.
}
\end{figure}

As before, individuals update their strategies based on probabilistic comparisons of their fitness with the fitness of other members of the population but now the pool of competitors is restricted to an individual's neighbourhood. This generates positive assortment among both strategic types through cluster formation, which benefits cooperators in social dilemmas by increasing interactions with other cooperators \citep{fletcher:PRSB:2009}. However, clusters can only expand along their periphery where cooperators are pitched against defectors to their disadvantage. In the Prisoner's Dilemma the benefits of assortment prevail and prevent the extinction of cooperation. The spatial arrangement reduces exploitation by defectors and hence enables cooperators and defectors to coexist \citep{nowak:Nature:1992b}. In contrast, in the Snowdrift game spatial structure may both enhance or inhibit cooperation \citep{hauert:Nature:2004} when compared to the equilibrium fraction in the replicator dynamics, see \fig{sdstruct}a.

When comparing interactions and updating in well-mixed and spatial games then the spatial structure obviously restricts the interaction partners of each individual to a small subset of the population but for pairwise comparison processes it also imposes that competing neighbours interact with each other at least with high probability. Consequently, it may be more appropriate to compare the effect of spatial structure to a well-mixed population where the two competitors interact as well. This corresponds to the fitness according to \eq{fifj}, which lowers the equilibrium frequency of cooperators in the Snowdrift game, \eq{xksd}. Using this scenario as a reference it turns out that spatial structure also invariably benefits cooperation, see \fig{sdstruct}b.

\subsubsection{Moran process}
Note, this relates to broader patterns observed in the spatial Moran process \citep{lieberman:Nature:2005}. The original Moran process \citep{moran:book:1962,nowak:Nature:2004} represents a stochastic birth-death process to model evolution in finite, well-mixed populations. Individuals are selected to reproduce with a probability proportional to their fitness. Their offspring inherits the strategic type of the parent and replaces a randomly selected member of the population. As a consequence the overall population size, $N$, remains constant and only the numbers of each strategic type changes. The dynamics in well-mixed populations is essentially indifferent to whether birth or death events are processed first \citep{zukewich:PlosOne:2013}. More specifically, it is intuitive that the differences are of order $1/N$ because the essential distinction between the birth-death and death-birth scenario is that the latter reduces the number of potential parents to $N-1$. In the limit of large populations, $N\to\infty$, the Moran process recovers the adjusted replicator dynamics \citep{traulsen:PRL:2005}. The adjusted replicator dynamics \citep{maynard-smith:book:1982} rescales time proportional to $1/\bar f$ (the inverse of the average population fitness) such that changes happen fast in poorly performing populations but slower in populations that are well off but exhibits the same equilibria as the (standard) replicator dynamics, \eq{rep}. 

\subsubsection{Birth-death and death-birth processes}
Interestingly, this is strikingly different in spatially structured populations: death-birth updates are significantly stronger promoters of cooperation than birth-death updates in the Prisoner's Dilemma. In particular, in the limit of weak selection, spatial structure provides \emph{no} support to cooperators for birth-death updating and cooperation remains doomed just as in well-mixed populations. However, for death-birth updating cooperators can thrive provided the benefits exceed the $k$-fold costs, $b>c k$, where $k$ denotes the (average) number of interaction partners of each individual, i.e. the degree of the graph \citep{ohtsuki:Nature:2006}. Intuitively, the reasons for this difference lie in the fact that for birth-death updates interaction partners tend to also compete with each other, whereas they tend not to for death-birth updates. In the first case this clearly puts cooperators at a disadvantage because they directly assist their competitors. For birth-death updating competition for reproduction is global and hence includes the interaction partners of a focal cooperator. In contrast, for death-birth updating competition is local because only the neighbours of a vacant site compete and they tend not to be neighbours themselves. The complementary conclusion is obtained if selection acts on death rather than birth, i.e. if individuals with a higher fitness live longer. Now the birth-death process supports cooperation because competition is local and the individuals competing for survival tend not to be interaction partners as well \citep{debarre:NatComm:2014}. More generally, what matters is the scale of competition or, more precisely, the overlap between competitors and interaction partners -- the more disjoint the two sets are, the better for cooperators.

\begin{figure}[tp]
\centerline{\includegraphics[width=0.8\linewidth]{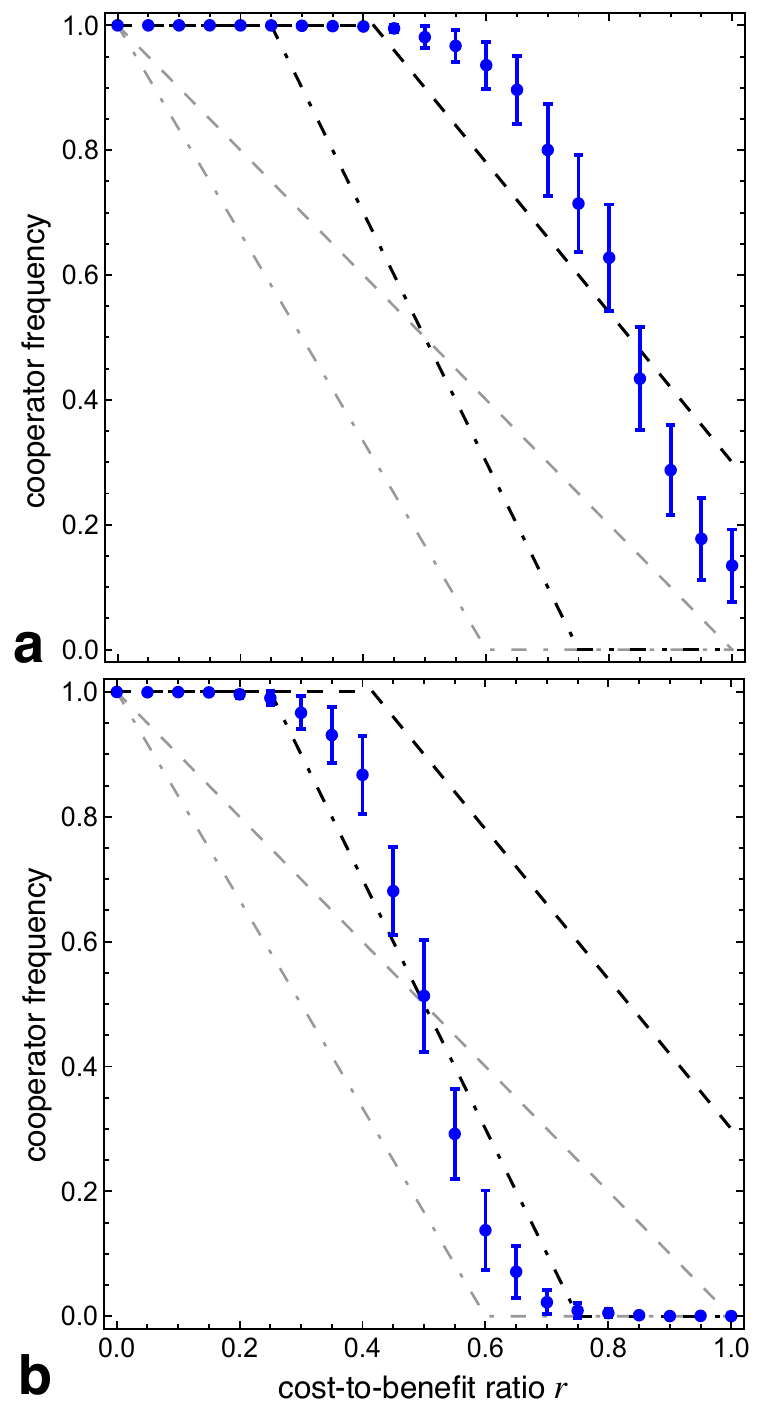}}
\caption{\label{fig:sdbddb}Equilibrium fraction of cooperators in the spatial Moran process for the Snowdrift game as a function of the cost-to-benefit ratio $r$ from individual based simulations (\textcolor{blue}{\small$\bullet$}, standard deviation as error bars) for \textbf{a} death-birth updating and \textbf{b} birth-death updating with selection strength $\omega=0.01$ on a $N=100\times100$ lattice with four nearest neighbours ($k=4$). The simulation results are averaged over $100$ runs each, relaxed over $5000$ generations and then taking the mean and standard deviation over the next $5000$ generations. The equilibria for the Moran process in well-mixed populations in the limit $N\to\infty$ are the same as for the replicator dynamics for standard sampling (grey dashed line). Accounting for the fact that competitors also interact for birth-death updating recovers modified sampling (grey dash-dotted line). For the spatial Moran process the equilibria are derived from pair approximation in the limit of weak selection, $\omega\ll1$, for birth-death updating (black dashed line) and death-birth updating (dash-dotted line).
}
\end{figure}

In the spatial Snowdrift game, the interplay between birth-death or death-birth updating and effects of population structure are more subtle, see \fig{sdbddb}.
Note that in the Moran process fluctuations are significantly larger than for pairwise comparison processes and hence absorption times are shorter. As a consequence, the equilibrium frequencies in simulations are averaged over multiple runs to reduce the effects of absorption in either homogeneous state of all cooperators or all defectors. Systematic deviations between analytical predictions in the weak selection limit and the simulations result from the finite selection strength as well as from the lattice structure. Deviations decrease for simulations on random regular graphs, which better reflect the assumptions of pair approximation.

For death-birth updating competitors tend not to interact but the details depend on the population structure. Two extremes are random regular graphs where neighbours do not interact in the limit $N\to\infty$ whereas on a lattice with Moore neighbourhood ($k=8$) some neighbours interact with up to half of the neighbourhood of the vacant site. For lattices with von Neumann neighbourhood ($k=4$) neighbours do not interact but remain close because they share half of their interaction partners. The case where competitors do not interact corresponds to the standard sampling in well-mixed populations, \eq{eqint}. Pair approximation and simulation results both confirm that spatial structure consistently increases the equilibrium frequency of cooperation, see \fig{sdbddb}a.

Conversely, for birth-death updating competitors do interact and hence corresponds to the modified sampling in well-mixed populations, \eq{eqintk}. Again, pair approximation and simulations confirms that spatial structure invariably increases cooperation, see \fig{sdbddb}b. However, as in \fig{sdstruct}b, comparisons to the standard sampling conclude that spatial structure is beneficial only at small $r$ and becomes detrimental at high $r$. The essential difference between birth-death and death-birth updating in structured populations can be captured in well-mixed populations by adjusting the sampling scheme to mimic the respective overlap between interaction partners and competitors.

\section{Random-matching models}
Random-matching models represents a well-mixed analogue to interactions on regular graphs with degree $k$ \citep{robson:JET:1996b,miekisz:JTB:2005}. In every time step individuals are arranged on a random regular graph, interact with their neighbours and a focal individual updates its strategy through a probabilistic payoff comparison with a randomly selected neighbour. Thus, the fitness is stochastic just as in structured populations but the redistribution of players between updates prevents clustering and destroys any spatial correlations while preserving all other aspects of the dynamical updating. This scheme naturally implements the modified sampling scheme, which includes interactions between focal and model individuals, and is therefore the default in random matching models.

\subsection{Infinite-population limit}
The composition of the population changes only if two individuals with different strategic types meet and probabilistically compare their payoffs. As before, let us focus on the simplest case of $2\times2$ games with two strategic types $A$ and $B$, \eq{2x2}. In infinite populations, the probability that two individuals of different type meet and compete is $x(1-x)$, where $x$ denotes the frequency of the $A$ type. For each individual this leaves an additional $k-1$ randomly chosen interaction partners to determine their respective fitness. The number of $A$ types among those follows a binomial distribution such that
\begin{align}
p(m) &= \binom{k-1}{m}x^{m}(1-x)^{k-1-m}
\end{align}
denotes the probability that $m$ of the $k-1$ co-players are of type $A$. Thus, the probability that a type $B$ individual imitates and switches to type $A$, or equivalently, gets displaced by offspring of the $A$ type, is given by
\begin{align}
p_{B\to A} &= \frac12+\frac{\omega}2\sum_{a=0}^{k-1} \sum_{b=0}^{k-1} p(a)p(b)\times\nonumber\\
	&\qquad\Big(a\,\alpha+(k-a)\beta-b\,\gamma+(k-b)\delta\Big)\\
	&=\frac12+\frac{\omega}2\Big(\beta-\gamma+(k-1)\times\nonumber\\
	&\qquad(\beta-\delta+x(\alpha-\beta-\gamma+\delta))\Big),
\end{align}
where the summations over $a, b$ indicate the number of $A$ types among the $k-1$ interaction partners of the focal $A$ and $B$ type, respectively. The converse probability, $p_{A\to B}$, reduces to
\begin{align}
p_{A\to B} &= \frac12-\frac{\omega}2\Big(\beta-\gamma+(k-1)\times\nonumber\\
	&\qquad(\beta-\delta+x(\alpha-\beta-\gamma+\delta))\Big),
\end{align}
such that $p_{A\to B}+p_{B\to A}=1$ holds \citep{molzon:DGA:2012}.

An interior stationary state of the random matching dynamics is, if it exists, given by $p_{A\to B}=p_{B\to A}$, such that the probability of an increase in $A$ types equals that of an increase in $B$ types. Because of $p_{A\to B}+p_{B\to A}=1$ both have to be $1/2$. The resulting equilibrium turns out to be the same as for the modified sampling, $x^\ast_k$, \eq{eqintk}. Decoupling interaction and competition of individuals through standard sampling shifts the interior stationary state to $x^\ast$, \eq{eqint}, provided it exists, and random matching recovers the equilibria of the standard replicator dynamics. If the interior stationary state is unstable, as in coordination games, or in the absence of an interior stationary state, the dynamics drives the population to either one of the absorbing homogeneous states with all $A$ or all $B$ types. Thus, in the infinite-population limit, both replicator dynamics and random-matching models give rise to the same long-run behaviour \citep{miekisz:JTB:2005}.

\subsection{Stochastic stability in finite populations}
In finite well-mixed populations of constant size, $N$, the state of the population is fully determined by the number of each strategic type and, more specifically, in the present case of $2\times2$ games by the number of $A$ types. A state is called \emph{stochastically stable} if it has a non-zero probability in the zero-noise limit of the stationary probability distribution \citep{foster:TPB:1990}. More precisely, this limit considers update rules where the strategy of individuals with higher fitness is always adopted (or, equivalently, the offspring of the fitter individual always succeeds in replacing the less fit) regardless of the magnitude of the fitness difference. Occasionally, with a small probability, $\epsilon$, an error happens and an inferior strategy is nevertheless adopted or a superior one is not. Note, we assume $\epsilon\ll 1/N$ to disentangle errors and probabilities for rare comparisons between different strategic types. For the pairwise comparison update, \eq{pji}, these assumptions correspond to the limit $\omega\to\infty$ while adding a noise term:
\begin{align}
\label{eq:pjiss}
p_{j\to i}=\epsilon+(1-2\epsilon)\Theta(f_i-f_j),
\end{align}
where $\Theta(x)$ denotes the Heaviside step function. With this updating the population eventually ends up in one of the absorbing homogeneous states with either all individuals of type $A$ or $B$. In order to prevent absorption, a small probability, $\mu$, is introduced with which mutations arise and an individual spontaneously switches to the opposite strategy. This turns the population dynamics into an ergodic process with a unique stationary probability distribution, which can interpreted as the fraction of time that the population spends in the corresponding state \citep{fudenberg:JET:2006}. For simplicity, we set $\mu=\epsilon$ and determine stochastic stability in the limit $\epsilon\to0$ for different sampling schemes, see \app{spd}. For dominance games, such as the Prisoner's Dilemma, stochastic stability and stability in infinite populations always coincide and do not depend on the sampling scheme. However, interesting differences arise for the Stag-hunt game, \eq{sh}, and the Snowdrift game, \eq{sd}, because of the bi-stable and co-existence dynamics, respectively.

\subsubsection{\label{sect:shrm}Stag-hunt game}
A crucial determinant of the evolutionary trajectory in the Stag-hunt game, \eq{sh}, is the basin of attraction of the two absorbing states of homogeneous cooperation or defection. 
The size of the two basins is determined by the location of the unstable interior fixed point, $x^\ast=a$ for standard sampling, and $x^\ast_k=a k/(k-1)$ 
for the modified sampling, \eq{shxk}. The strategy with the larger basin of attraction is called risk-dominant. For \eq{sh} and standard sampling this means that $C$ is risk-dominant for $a<1/2$ while $D$ is risk-dominant for $a>1/2$. Similarly, $C$ is risk-dominant for $a<(k-1)/(2k)$ for the modified sampling and $D$ is risk-dominant for the reverse inequality. 

In the replicator dynamics, the evolutionary end state is simply determined by the chance with which the initial configuration falls into one or the other basin of attraction. 
In contrast, for the stochastic dynamics in finite populations, the only stochastically stable state is a homogeneous population 
with all individuals adopting the risk-dominant strategy \citep{kandori:ECO:1993,miekisz:LNM:2008}. 

\begin{figure}[tp]
\centerline{\includegraphics[width=0.8\linewidth]{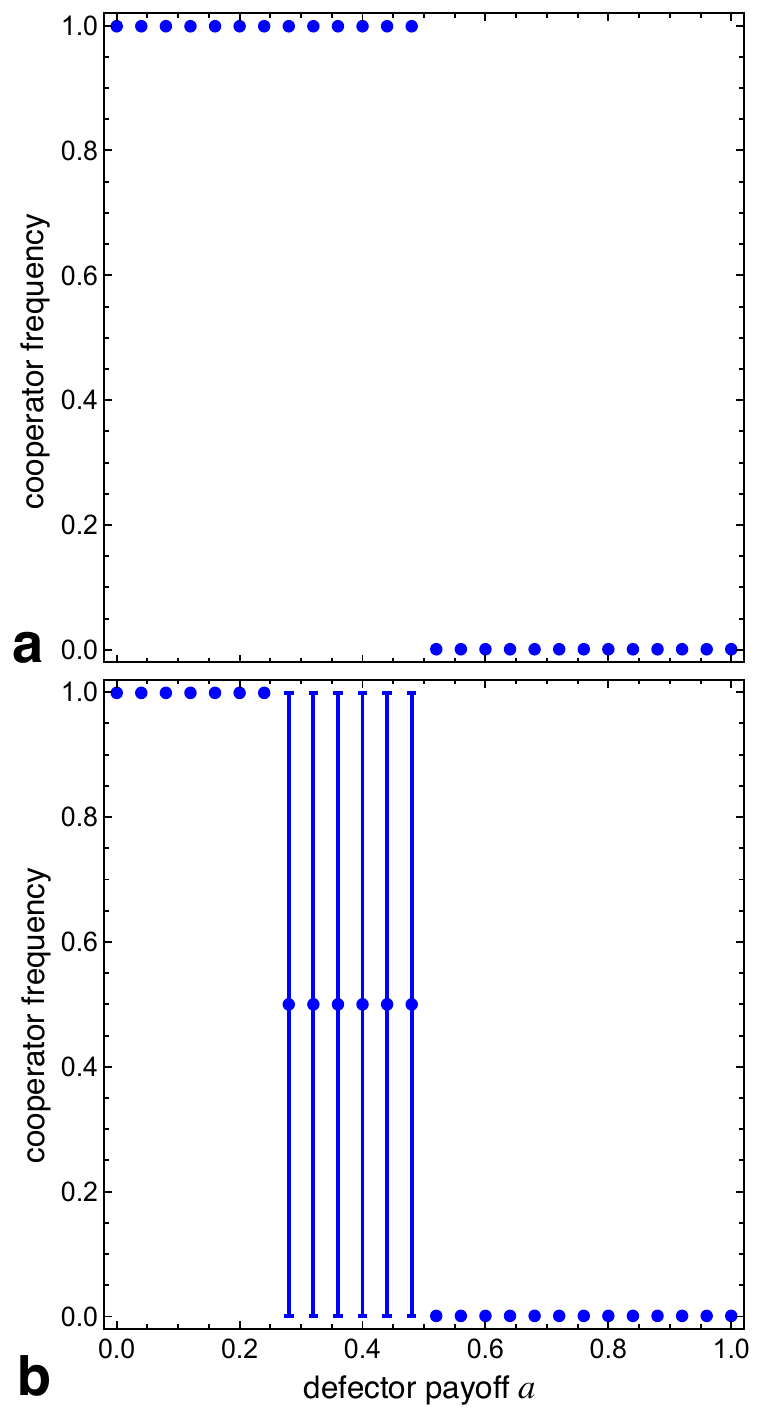}}
\caption{\label{fig:shss}Stochastic stability in the Stag-hunt game for random matching in finite populations ($N=100$) for \textbf{\textsf{a}} standard sampling 
and \textbf{\textsf{b}} modified sampling (focal and model individuals always interact) with $k=4$ interaction partners and $\epsilon=0.001$ as a function of the payoff for a hare, 
$a$, see \eq{sh} (\textcolor{blue}{\small$\bullet$}, mean of stationary distribution with standard deviation as error bars). \textbf{\textsf{a}} for standard sampling, 
the risk dominant strategy is stochastically stable: for $a<1/2$ cooperation is risk dominant whereas for $a>1/2$ defection is risk dominant. 
\textbf{\textsf{b}} In contrast, for the modified sampling, cooperation is stochastically stable only for $a<1/4$, whereas defection remains the only stable state for $a>1/2$. 
However, for $1/4<a<1/2$ both homogeneous states are stochastically stable and attract the same probability mass, which results in a mean of $1/2$.}
\end{figure}

Intuitively, stochastic stability is determined by the number of rare events, i.e. transitions with probability $o(\epsilon)$, 
that are required to switch from one homogeneous state to the other, stochastic stability requires $\epsilon\ll 1/N$ because otherwise many more transitions become relevant. Naturally, the one requiring fewer rare events to reach dominates 
and is therefore stochastically stable, see \app{spd}. More specifically, for the Stag-hunt game we obtain the following theorem:
\begin{theorem}
In the Stag-hunt game, \eq{sh}, the homogeneous state with all cooperators, $C$, is stochastically stable under random matching if 
\[a<\max_{i=1,\ldots, k}\min\{\frac{i-1}{k},1-\frac{i}{k}\},\]
while the homogeneous state with all defectors, $D$, is stochastically stable if  
\[a>\min_{i=1,\ldots, k}\max\{\frac{i-1}{k},1-\frac{i}{k}\}.\]
\end{theorem}
The proof is provided in \app{sssh}. Note that for even $k$, the second inequality reduces to $a>1/2$. From the theorem follows that $C$ is stochastically stable, for example, 
for $k=4$ if $a<1/4$ and $D$ is stochastically stable if $a>1/2$, while for $1/4 \leq a \leq 1/2$ the stationary probability mass is split among the $C$ and $D$ states 
and hence both are stochastically stable, see \fig{shss}b.

In order to highlight the effects of sampling, let us now consider a random-matching model without interactions between focal and model individuals by randomly choosing the model 
from the entire population and not just from interaction partners of the focal individual. Note that this change does not affect the derivation of the stochastic payoffs. 
Interestingly, changing the sampling leaves the conditions for stochastic stability unchanged but this is not reflected in numerical results, see \fig{shss}a. 
It turns out that the dynamics in the region $1/4<a<1/2$ for $k=4$ is more subtle and is not only constrained by rare events based on the error probability $\epsilon$ 
but also by events of the order $1/N^2$ and $1/N^3$.

In the Stag-hunt game, the rare type is always at a disadvantage and hence the probability that the number of $C$-players increases from one to two, $1\to2$, is of order $o(\epsilon)$ regardless of $a$. Now consider the transition $2\to3$ with $k=4$: in the best scenario for $C$, a $C$-player interacts with the other $C$-player (plus $3$ $D$-players) and receives a payoff of $1$ while competes with a $D$-player who gets a payoff of $4a$ regardless of its interaction partners. Thus, for $a>1/4$ the transition remains of order $o(\epsilon)$. More interesting is the transition $3\to4$: again, in the best scenario a $C$-player interacts with both $C$-players (plus $2$ $D$-players) and receives a payoff of $2$ such that now the transition is viable for $a<1/2$. However, this best scenario occurs only with probability of the order $1/N^2$ and hence makes a significant contribution only if $\epsilon<1/N^2$.

Analogous arguments apply if $D$-players are rare and $C$-players common. The transition from a single $D$ to two $D$'s, or $N-1\to N-2$ $C$-players, and from $N-2\to N-3$ are viable for $a>3/4$ and $a>1/2$, respectively, in the best scenario for the $D$-player. More interestingly, the transition $N-3\to N-4$ is viable for $a>1/4$, in the best scenario but this occurs only with probability of the order $1/N^3$ based on the probability that the $C$ competitor has interacted with all $D$-players in the population.

Comparing the transitions $3\to4$ and $N-3\to N-4$ $C$-players, we note that for $1/4<a<1/2$, the former transition is more likely by a factor of $1/N$ and hence for small $\epsilon$ and large $N$ (more precisely, $\epsilon>1/N^2$), almost all probability mass is concentrated at $C$ even though $D$ is, in principle, stochastically stable too. For larger $N$, with $\epsilon<1/N^2$, the result remains unchanged but  all other transition probabilities have to be taken into account, too.

This is easily generalized to arbitrary $k$: for a low but fixed $\epsilon$ and sufficiently large $N$, the stationary probability distribution assigns almost all mass to $C$ for $a<\min_{i=1,\ldots, k}\max\{\frac{i-1}{k},1-\frac{i}{k}\}$ and to $D$ otherwise.

\subsubsection{\label{sect:sdrm}Snowdrift game} 
\begin{figure}[tp]
\centerline{\includegraphics[width=0.8\linewidth]{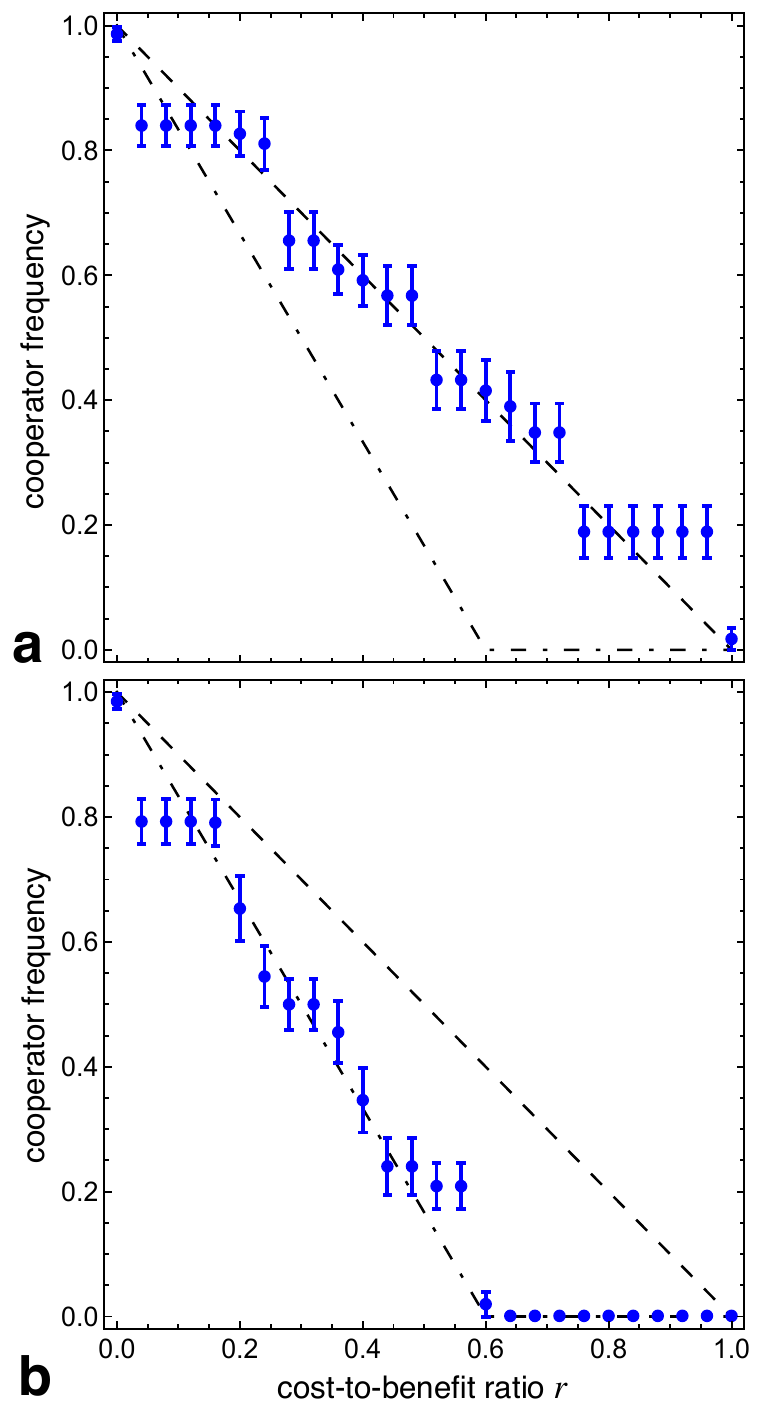}}
\caption{\label{fig:sdss}Stochastic stability in the Snowdrift game for random matching in finite populations ($N=100$) for \textbf{\textsf{a}} standard sampling and \textbf{\textsf{b}} modified sampling (focal and model individuals always interact) with $k=4$ interaction partners and $\epsilon=0.001$ as a function of the cost-to-benefit-ratio, $r$, see \eq{sd} (\textcolor{blue}{\small$\bullet$}, mean of stationary distribution with standard deviation as error bars). \textbf{\textsf{a}} for standard sampling the interior equilibrium $x^\ast$ is always stochastically stable. Note, $x^\ast$ coincides with the homogeneous equilibria for $r=0$ and $r=1$. \textbf{\textsf{b}} In contrast, for the modified sampling, defection is stochastically stable for $r>3/5$, whereas the interior equilibrium remains stable for $r<3/5$.
}
\end{figure}

The replicator dynamics admits a single stable equilibrium where cooperators and defectors co-exist in the Snowdrift game, \eq{sd}. 
The equilibrium fraction of cooperators depends on the sampling scheme and is lower (or disappears) if competitors also interact, c.f. \eq{xksd} and \fig{sdmix}.
However, in finite populations, stochastic fluctuations inevitably drive the population to either one of the homogeneous states, which are absorbing in the absence of mutations. 
In the Snowdrift game, absorption times scale exponentially with population size $N$ \citep{antal:BMB:2006}. Thus, for small mutation rates, $\mu$, 
the splitting of the probability mass of the stationary distribution between the two homogeneous states and the co-existence state is non-trivial 
but remains accessible for stochastic stability.
\begin{theorem}
In Snowdrift games, \eq{sd}, the homogeneous state with all defectors, $D$, is stochastically stable under random matching if 
\[r>(k-1)/(k+1),\]
while for $r<(k-1)/(k+1)$, the stationary distribution assigns non-zero probability to interior states. 
\end{theorem}
The proof is provided in \app{sssd}. Interestingly this implies that $C$ is never stochastically stable. For $k=4$, it follows that $D$ is stochastically stable for $r>3/5$, see \fig{sdss}b.
The discontinuous jumps in the frequency of cooperators can be associated with thresholds of $r$ that ensure that $C$-players outcompete $D$-players in relevant configurations. Naturally, which configurations are relevant depends on the frequency of cooperators. For example, if cooperators are rare, most likely neither $C$-players nor $D$-players interact with another $C$ and thus their respective payoffs are $k(b-c)$ and $b$ (recall that by default competitors also interact in random matching models). The $C$-players win if $r<(k-1)/(k+1)$, which corresponds to the stochastic stability threshold above which the homogeneous $D$ state is stable. At slightly higher frequencies of cooperators both $C$-players and their competing $D$-players are both likely to interact with one randomly chosen $C$-player, which yields $b-c/2+(k-1)(b-c)$ for the $C$-player and $2b$ for the $D$-player. Thus, the $C$-player wins if $r<(k-2)/(k+1)$. For still higher frequencies of cooperators, both competitors likely interact with two random $C$-players, which translates into a thresholds of $r<(k-3)/(k+1)$ for cooperators to win. Thus we expect $k-1$ transition points where the stochastic sampling of interaction partners favours and reinforces particular frequencies of cooperators and relates to the three jumps observed for $k=4$ in \fig{sdss}b at $r\approx0.2, 0.4, 0.6$. The above argument also suggests that the plateaus should occur in increments of $1/k$, which also matches \fig{sdss}b but naturally the plateaus are less pronounced for frequencies around $50\%$ because many other types of matchings are likely and introduce further and finer grains of thresholds.

In a random matching model with standard sampling, i.e. in the case where focal and model individuals do not necessarily interact, the results are again quite different. In fact, for $0<r<1$ neither the homogeneous $C$ nor $D$ state is stochastically stable and effectively all probability mass is assigned to interior states. The Snowdrift game favours rare types, which means 
that the transitions $1\to2$ and $N-1\to N-2$ of the number of $C$-players are highly likely regardless of $r$. More precisely, a single $C$-player in the population gets a payoff of $k(b-c)$and, in the worst case, competes with a $D$-player who also happened to interact with the $C$ and thus has a payoff of $b$. Consequently the $C$-player loses if $r>(k-1)/(k+1)$. However, this scenario occurs only with a probability of the order of $1/N$ and for smaller $r$ or other matchings the $C$-player always wins. Conversely, a single $D$-player in the population wins with certainty regardless of $r$: even in the worst case the $D$-player gets a payoff of $k\cdot b$ and outcompetes $C$-players with a payoff of $k(b-c)$ from interactions with other $C$'s. As a consequence, although $D$ remains stochastically stable, for fixed $\epsilon$ and large $N$, the stationary probability assigns almost all probability mass to interior states, \fig{sdss}a.

The frequency of cooperators again exhibits discontinuous jumps that relate to thresholds of $r$, which enables cooperators to outcompete defectors in relevant configurations. Naturally, both the thresholds and the levels of the plateaus are different because competing $C$-players and $D$-players no longer necessarily interact. If both interact with one $C$-player they get $b-c/2+(k-1)(b-c)$ and $b$, respectively, such that $C$-players win if $r<(k-1)/k$. Similarly, if both interact with two $C$-players, the $C$-player wins if $r<(k-2)/k$. Thus, we again expect $k-1$ transition points that favour certain frequencies of cooperators and are observed for $k=4$ in \fig{sdss}a at $r\approx0.25, 0.5, 0.75$. In contrast to \fig{sdss}b the increment between plateau levels seems closer to $1/(k+1)$ but such differences are not surprising because these levels are the result of dynamical feedback between sampling probabilities and transition probabilities.

\section{Discussion}
For modelling evolutionary trajectories, the sampling process, which determines the interaction partners and hence the fitness of competing individuals, has a decisive impact on the evolutionary outcome. Naturally, the sampling is affected by the structure of the population (or lack thereof) but another crucial aspect is whether or not competing individuals also interact.

In a microscopic interpretation of the canonical dynamics of evolutionary game theory, the replicator equation \citep{hofbauer:CUP:1998}, interaction partners are randomly sampled and competing individuals do not interact. More precisely, in unstructured finite populations of size $N$ the chance that the focal individual and its competitor interact is of order $1/N$ and hence vanishes in the infinite population limit of the replicator dynamics. Modifying the sampling scheme to always include interactions between competitors changes the dynamics quantitatively and, at times, even qualitatively in co-existence and coordination games. In social dilemmas, the modified sampling scheme invariably favours defectors. While this merely speeds up the demise of cooperators in the Prisoner's Dilemma, it shifts the interior equilibrium in the Snowdrift game and reduces the equilibrium fraction of cooperators and similarly increases the basin of attraction of defectors in the Stag-hunt game. In the latter two cases the change in sampling may even eliminate the interior fixed point altogether turning defection into a dominant strategy, just as in the Prisoner's Dilemma. In particular, if individuals exclusively interact with their competitor, neither co-existence nor bi-stability is possible and one type invariably dominates the other.

In social dilemmas, assortment promotes cooperation by reducing exploitation by defectors, for example, through cluster formation in structured populations. However, this effect is reduced if competitors also interact, which is the case in traditional models with identical interaction and reproduction graphs \citep{nowak:Nature:1992b,hauert:Nature:2004,ohtsuki:Nature:2006} (some notable exceptions are \citep{ifti:JTB:2004,ohtsuki:PRL:2007,debarre:NatComm:2014}). Even if interaction partners are randomly sampled from an individual's limited local neighbourhood, chances are high that this includes any neighbour who is challenging the focal individual.



In the Prisoner's Dilemma the benefits of cluster formation tend to prevail and spatial structure is capable of supporting cooperation \citep{nowak:Nature:1992b}. This also applies to the spatial Moran process except in the weak selection limit of birth-death updating where the benefits of assortment are cancelled by competing with interaction partners \citep{ohtsuki:Nature:2006,debarre:NatComm:2014}. In the Stag-hunt game spatial structure in general simply favours the risk-dominant strategy \citep{hauert:IJBC:2002} but adding noise can result in surprising qualitative changes \citep{szabo:PRE:2016}. Interestingly, the situation is more subtle for the spatial Snowdrift game \citep{hauert:Nature:2004} and depends on the reference scenario. 
When compared to the replicator dynamics, the effects of space can inhibit or enhance cooperation depending on the game parameters. 
However, 
when compared to the equilibrium fraction of cooperators 
for the replicator dynamics with modified sampling, then spatial structure again has an invariably beneficial impact on cooperation also in the Snowdrift game. 

Random matching models provide another interesting alternative to identify and quantify effects of population structure because they assume an underlying population structure but then randomly reshuffle individuals between updates to prevent spatial correlations. Thus the default sampling of interaction partners includes competitors for random matching but of course the sampling scheme can easily be adjusted to reflect the standard sampling of the replicator dynamics. In the limit of infinite populations, random matching recovers the long-term dynamics of the replicator equation for both sampling types. In finite populations the evolutionary outcome is determined by stochastic stability, which identifies all states of the population with a non-zero allocation of probability mass in the zero-noise limit. The results agree well with the expectations based on the replicator equation with modified or standard sampling. Interestingly, however, confirming stochastic stability through simulations and numerical analysis turns out to be surprisingly challenging, which highlights that this zero-noise limit is fickle and may be of limited relevance for modelling purposes.

The differences in the dynamics based on whether interactions include competitors is mirrored in the differences between birth-death and death-birth updating in the spatial Moran process \citep{ohtsuki:Nature:2006,debarre:NatComm:2014}. For example, in the weak-selection limit, death-birth updating can sustain cooperation in the Prisoner's Dilemma if $b>c\cdot k$, while birth-death updating cannot. The reason for this difference is that for death-birth updating, neighbours of a vacant site compete to repopulate the vacancy but neighbours tend not to be neighbours themselves and hence tend not to interact with each other. Conversely, for birth-death updating, all individuals compete for reproduction and, in particular, each individual also competes with its interaction partners and consequently, cooperators provide direct support to competitors at a disadvantage to themselves.

The upshot is that for the success of cooperation it is crucial that individuals carefully sample their interaction partners not only to reduce exploitation by defectors but also to avoid interactions with competitors.

\acknowledgements{JM would like to thank the National Science Centre (Poland) for a financial support under Grant No. 2015/17/B/ST1/00693 
and CH acknowledges a financial support from the Natural Sciences and Engineering Research Council of Canada (NSERC), grant RGPIN-2015-05795.}

\appendix
\section{\label{app:spd}Stationary probability distributions of ergodic Markov chains}
We assume that the system follows some deterministic rule with the probability $1-\epsilon$ and with the probability $\epsilon$, 
a mistake is made that moves the system in the other direction. 

Let $(\Omega,P_{\epsilon})$ be a discrete-time Markov chain with a finite state space $\Omega$ and transition probabilities given by $P_{\epsilon}: \Omega \times \Omega \rightarrow [0,1]$.
We assume that one can get with non-zero probability from any state to any other state in a finite number of steps. It follows that the Markov chain has the unique stationary probability
distribution $\mu_{\epsilon}$. We say that the state $x \in \Omega$ is stochastically stable if it has a non-zero probability in the stationary probability distribution, that is 
\begin{equation}
\lim_{\epsilon \rightarrow 0} \mu_{\epsilon}(x) > 0.
\end{equation}
It means that the state $x$ is observed in the long run with a non-zero frequency. We are usually interested in situations where this frequency is equal to $1$.

For $x \in \Omega$, an $x$-tree with the root at $x$ is a directed graph on $\Omega$ (connecting all vertices) such that from every $y \neq x$ there is a unique path to $x$ 
and there are no outgoing edges out of $x$. Denote by $T(x)$ the set of all $x$-trees and let 
\begin{equation}
q_{\epsilon}(x)=\sum_{d \in T(x)} \prod_{(y,y') \in d}P_{\epsilon}(y,y'), 
\end{equation}
where $P_{\epsilon}(y,y')$ is the element of the transition matrix (that is, a conditional probability that the system will be at the state $y'$ at time $t$ provided it was at the state $y$
at time $t-1$) and the above product is with respect to all edges of the $x$-tree $d$. In our finite population models, $P_{\epsilon}(y,y')$ is a polynomial in $1-\epsilon$ 
and $\epsilon$ or it is equal to $0$.

Now one can show that (the Tree Lemma) \citep{freidlin:book:1984,shubert:IEEE:1975}
\begin{equation}\label{eq:statstate}
\mu_{\epsilon}(x)=\frac{q_{\epsilon}(x)}{\sum_{y \in \Omega}q_{\epsilon}(y)}
\end{equation}
for all $x \in \Omega.$

A state is absorbing if it attracts nearby states in the zero-noise (no mistakes) dynamics ($\epsilon=0$). That is after a finite number of steps of the zero-noise dynamics, 
the system arrives at one of the absorbing states and stays there forever. It follows from \eq{statstate} that the stationary probability distribution can be written 
as the ratio of two polynomials in $\epsilon$. Hence any non-absorbing state has zero probability in the stationary distribution in the zero-noise limit ($\epsilon \rightarrow 0$). 
Moreover, in order to study the zero-noise limit of the stationary distribution, it is enough to consider paths between absorbing states. Let us assume that the system has two absorbing
states. Let $m_{xy}$ be a minimal number of mistakes needed to make a transition from the state $x$ to $y$ and $m_{yx}$ the minimal number of mistakes to make a transition from $y$ to $x$.
Then $q_{\epsilon}(x)$ is of the order $\epsilon^{m_{yx}}$ and $q_{\epsilon}(y)$ is of the order $\epsilon^{m_{xy}}$. If for example, $m_{yx}<m_{xy}$, then it follows 
that $\lim_{\epsilon \rightarrow 0}\mu_{\epsilon}(x) = 1$, hence $x$ is stochastically stable.

In one-dimensional models considered here (biased random walk on $N+1$ integers with absorbing states at $0$ and $N$), the situation is much simpler -- all states have unique trees and the only absorbing states are $0$ and $N$. The unique tree of an absorbing state is given by a direct path from the other absorbing state, while for any interior state, its tree is made of two directed paths from $0$ and $N$.

\section{\label{app:sssh}Stochastic stability in the Stag-hunt game - Proof of Theorem 1}
First we find the threshold for the extinction for $D$, that is the biggest $a$ such that $C$ is stochastically stable. 
The best scenario for $C$, for transitions $i \to i+1, \; i=1,...,k$, is that a model $C$-player interacts with $i-1$ $C$-players and one $D$-player. 
Then for $a < (i-1)/k, \; 1 < i \leq k$, the transition $i \to i+1$ takes place with the probability $1-\epsilon$. For $i=N-1,...,N-k$, the best scenario for $D$, 
is that a model $C$-player interacts with $N-i$ $D$-players. Then for $a < 1-i/k, \;  N-k \geq i < N$, transitions $i \to i+1$ take place with the probability $1-\epsilon$.
For both inequalities for $a$ to hold we take a minimum value of two bounds, $\min\{\frac{i-1}{k},1-\frac{i}{k}\}$. To get a threshold for the extinction of $D$ 
we take a maximum value of the above expression with respect to $i = 1,...,k$.

Now we find the threshold for the extinction for $C$, that is the smallest $a$ such that $D$ is stochastically stable. 
If $a > (i-1)/k, \; 1 < i \leq k$, then even for the best scenario for $C$, transitions $i \to i+1$ take place with the probability $1-\epsilon$.
For $a > 1-i/k, \;  N-k \geq i < N$, for the best scenario for $D$ transitions $i \to i-1$ takes place with the probability $1-\epsilon$.
For both inequalities for $a$ to hold we take a maximum value of two bounds, $\max\{\frac{i-1}{k},1-\frac{i}{k}\}$. To get a threshold for the extinction of $C$ 
we take a minimum value of the above expression with respect to $i = 1,...,k$.

\section{\label{app:sssd} Stochastic stability in the Snowdrift game - Proof of Theorem 2}
For $z=N-1$, the payoff of the only $D$-player is bigger than the payoff of any $C$-player. Therefore the population moves from $z=N-1$ to $z=N-2$ with probability $1-\epsilon$.
If $k(b-c)<b$, that is $r>(k-1)/(k+1)$, then the payoff of a single $C$-player is smaller than the payoff of any $D$-player and therefore, 
the population moves from $z=1$ to $z=0$ with probability $1-\epsilon$. Hence the tree of $z=0$ is of order $\epsilon$ and both $z=N$ and $z=x^\ast$ have trees of order $\epsilon^{3}$. It follows from the Tree lemma (see \app{spd}) that the homogeneous population with all defectors is stochastically stable.

Conversely, for $r<(k-1)/(k+1)$, all three states $z=0$, $z=N$, and $z=x^\ast$, have trees of order $\epsilon^{2}$. It follows that in the zero-noise limit, the stationary distribution assigns non-zero probability to interior states with $C$ and $D$ strategies co-existing. 

\bibliography{ET}

\end{document}